\documentclass[aps,eprint,twocolumn,longbibliography,superscriptaddress, amsmath,amssymb]{revtex4-2}
\usepackage{times}
\usepackage{graphicx}
\usepackage{float}
\usepackage{color}
\usepackage{subfigure}
\usepackage{epstopdf}
\usepackage[colorlinks=true,linkcolor=blue,citecolor=blue]{hyperref}
\usepackage[normalem]{ulem}
\usepackage{mathrsfs}
\usepackage{amssymb}
\usepackage{algpseudocode}
\usepackage{algorithm}
\usepackage{amsmath}
\usepackage{braket}
\usepackage{booktabs}
\usepackage{chngcntr}
\usepackage{xspace}
\usepackage{textcomp}
\usepackage{textcase}
\usepackage{setspace}
\usepackage[T1]{fontenc}
\usepackage{float}
\usepackage{graphicx}
\usepackage{multirow}
\usepackage{mathtools}
\usepackage{threeparttable}
\usepackage[vlined, ruled, algo2e, linesnumbered]{algorithm2e}
\usepackage{changes}
\usepackage{makecell}

\begin{document}

\title{Successive magnetic transitions in the spin-5/2 easy-axis triangular-lattice antiferromagnet Na$_2$BaMn(PO$_4$)$_2$: A neutron diffraction study}

\author{Chuandi Zhang}
\affiliation{School of Physics, Beihang University, Beijing 100191, China}

\author{Junsen Xiang}
\affiliation{Beijing National Laboratory for Condensed Matter Physics, Institute of Physics, Chinese Academy of Sciences, 100190 Beijing, China}

\author{Cheng Su}
\affiliation{School of Physics, Beihang University, Beijing 100191, China}

\author{Denis Sheptyakov}
\affiliation{Laboratory for Neutron Scattering and Imaging, Paul Scherrer Institut, CH-5232 Villigen-PSI, Switzerland}

\author{Xinyang Liu}
\affiliation{School of Physics, Beihang University, Beijing 100191, China}

\author{Yuan Gao}
\affiliation{School of Physics, Beihang University, Beijing 100191, China}

\author{Peijie Sun}
\affiliation{Beijing National Laboratory for Condensed Matter Physics, Institute of Physics, Chinese Academy of Sciences, 100190 Beijing, China}

\author{Wei Li}
\affiliation{CAS Key Laboratory of Theoretical Physics, Institute of Theoretical Physics, Chinese Academy of Sciences, Beijing 100190, China}
\affiliation{CAS Center for Excellence in Topological Quantum Computation, University of Chinese Academy of Sciences, Beijng 100190, China}
\affiliation{Peng Huanwu Collaborative Center for Research and Education, Beihang University, Beijing 100191, China}

\author{Gang Su}
\affiliation{Kavli Institute for Theoretical Sciences, and School of Physical Sciences, University of Chinese Academy of Sciences, Beijing 100049, China}
\affiliation{CAS Center for Excellence in Topological Quantum Computation, University of Chinese Academy of Sciences, Beijng 100190, China}

\author{Wentao Jin}
\email{wtjin@buaa.edu.cn}
\affiliation{School of Physics, Beihang University, Beijing 100191, China}

\date{\today}

\begin{abstract}
Motivated by the recent observations of various exotic quantum states in the equilateral triangular-lattice phosphates Na$_2$BaCo(PO$_4$)$_2$ with $J\rm_{eff}$ = 1/2 and Na$_2$BaNi(PO$_4$)$_2$ with $S$ = 1, the magnetic properties of spin-5/2 antiferromagnet Na$_2$BaMn(PO$_4$)$_2$, their classical counterpart,  are comprehensively investigated experimentally. DC magnetization and specific heat measurements on polycrystalline samples indicate two successive magnetic transitions at $T\rm_{N1}$ $\approx$ 1.13 K and $T\rm_{N2}$ $\approx$ 1.28 K, respectively. Zero-field neutron powder diffraction measurement at 67 mK reveals a Y-like spin configuration as its ground-state magnetic structure, with both the $ab$-plane and $c$-axis components of the Mn$^{2+}$ moments long-range ordered. The incommensurate magnetic propagation vector $k$ shows a dramatic change for the intermediate phase between $T\rm_{N1}$ and $T\rm_{N2}$, in which the spin state is speculated to change into a collinear structure with only the $c$-axis moments ordered, as stabilized by thermal fluctuations. The successive magnetic transitions observed in Na$_2$BaMn(PO$_4$)$_2$ are in line with the expectation for a triangle-lattice antiferromagnet with an easy-axis magnetic anisotropy. 

\end{abstract}

\maketitle

\section{Introduction}
Triangular-lattice antiferromagnets with strong geometrical frustration have attracted tremendous research interest in the past few decades \cite{Ramirez1994, Collins1997, Kanoda2011}, due to
the possibility to realize a novel quantum spin liquid state \cite{Anderson1973, Balents2010, Zhou2017, Broholm2020, Wen2019} and various exotic spin states \cite{Chubukov1991, Starykh2015, Khatua2023}. Among them, equilateral triangular-lattice transition-metal phosphates Na$_2$Ba$T$(PO$_4$)$_2$ ($T$ = Co, Ni, Mn) with the glaserite-type structure have attracted tremendous attention in the past few years, due to the recent observations of a giant magnetocaloric effect associated with the intriguing spin supersolidity in Na$_2$BaCo(PO$_4$)$_2$ (NBCP) \cite{gao2022, xiang2024}, Bose-Einstein condensation of the two-magnon bound state in Na$_2$BaNi(PO$_4$)$_2$ (NBNP) \cite{sheng2023}, and multiple field-induced quantum spin state transitions in this family \cite{li2020, Sheng2022, li2021, kim2022}. 

NBCP with an effective spin $J\rm_{eff}$ = 1/2, that firstly proposed as a quantum spin liquid candidate \cite{zhong2019, lee2021, Wellm2021}, was revealed to be a rare and nearly ideal material realization of the spin-1/2 easy-axis XXZ model, for which a Y-like spin supersolid state is realized as its magnetic ground state in zero field below $T\rm_{N}$ $\sim$ 0.15 K \cite{li2020, gao2022, xiang2024, Sheng2022}. On the other hand, NBNP with $S$ = 1, whose spin Hamiltonian was found to be dominated by nearest-neighbor XXZ-type exchange interactions plus a single-ion anisotropic term, is speculated to exhibit a spin nematic supersolid ground state below $T\rm_{N}$ $\sim$ 0.43 K \cite{sheng2023}. In both compounds, the strong quantum spin fluctuations play a decisive role in realizing those exotic quantum states. Therefore, it is worthwhile to investigate their classical counterpart with a much larger spin quantum number for comparison. 

Na$_2$BaMn(PO$_4$)$_2$ (NBMP), a mineral discovered and named as ``iwateite'' in 2014 \cite{Iwateite2014}, consists of  equilateral triangular lattice of spin-5/2 Mn$^{2+}$ ions. Although structural investigations from different groups using x-ray diffraction (XRD) yield controversial conclusions about its space group ($\mathit{P}$$\bar{3}$ vs. $\mathit{P}$$\bar{3}m1$) \cite{Iwateite2014, Nenert2020, kim2022}, our recent room-temperature neutron powder diffraction (NPD) measurement indicates the elimination of the mirror symmetry due to the displacements of oxygen atoms and confirms its true space group as $\mathit{P}$$\bar{3}$ \cite{zhang2024a}, well consistent with an independent NPD study from Kajita $et$ $al.$ \cite{Kajita2024}. On the other hand, the magnetic properties of this compound are still largely unexplored, to the best of our knowledge. Zero-field thermodynamic measurement on a single-crystal sample and dc magnetization measurement on a polycrystalline sample consistently reveal two magnetic transitions in NBMP at $T\rm_{N1}$ $\approx$ 1.15 K and $T\rm_{N2}$ $\approx$ 1.30 K \cite{kim2022, zhang2024a}, respectively, whose nature is not clearly understood yet. When a magnetic field is applied at 47 mK along the $c$ axis of the single-crystal sample, multiple field-induced spin states including the up-up-down (UUD) phase on the 1/3 magnetization plateau were disclosed by detailed ac magnetic susceptibility measurements, suggesting an easy-axis magnetic anisotropy \cite{kim2022}. Based on the assumption of a classical two-dimensional (2D) triangular-lattice Heisenberg antiferromagnetic model, a Y state was expected as the magnetic ground state of NBMP, awaiting the experimental verification using microscopic magnetic probes such as neutron diffraction.

In this paper, we have synthesized high-purity polycrystalline samples of NBMP and conducted temperature-dependent NPD experiments. The Y state is confirmed to be the magnetic ground state of NBMP, similar to the case of NBCP spin supersolid. With increasing temperature, the magnetic propagation vector $k$ shows a dramatic change for $T\rm_{N1}$ $\textless$ $T$  $\textless$ $T\rm_{N2}$ and the refinement to the diffraction pattern supports a $c$-axis collinear spin structure for this intermediate phase. The occurrence of two successive magnetic transitions is consistent with the expectation for a triangular-lattice antiferromagnet with an easy-axis anisotropy, with the $c$-axis and $ab$-plane components of the Mn$^{2+}$ moments ordered separately at $T\rm_{N2}$ and $T\rm_{N1}$. $k_z$, the out-of-plane component of the magnetic propagation vector $k$, is incommensurate in both two phases, suggesting significant interlayer couplings between the triangular Mn layers. 

\section{Methods}

Polycrystalline samples of NBMP were synthesized by a standard solid-state reaction method. 
Stoichiometric amounts of dried Na$_2$CO$_3$(99.99$\%$), BaCO$_3$ (99.99$\%$), MnO (99.5$\%$) and (NH$_4$)$_2$HPO$_4$ (99.99$\%$) were mixed and thoroughly ground with the catalyst NH$_4$Cl (99.99$\%$), in the molar ratio of 2:1. The mixture were pelletized, transferred into alumina crucibles, sintered at 850 $^\circ$C for 24~h in air, and cooled down to room temperature. The sintering process was repeated several times to minimize the impurities. The phase purity of the polycrystalline samples was checked by XRD measurement, utilizing a Bruker D8 ADVANCE x-ray diffractometer with Cu-K$\alpha$ radiation ($\lambda$  = 1.5406 \AA). 

DC magnetization and specific heat of the sample were measured using a Quantum Design Magnetic Property Measurement System (MPMS) and Physical Property Measurement System (PPMS), respectively, equipped with the $^{3}$He insert. 

The NPD experiments were performed on the High Resolution Powder diffractometer for Thermal neutrons (HRPT) ~\cite{fischer2000} at the Swiss Neutron Spallation Source (SINQ), at the Paul Scherrer Institute in Villigen, Switzerland. Polycrystalline samples of NBMP with a total mass of 9.7 grams were loaded into a cylindrical copper cell, and cooled down to the base temperature of 67 mK using a dilution insert inside a top-loading cryostat. The diffraction patterns were collected using wavelengths of $\lambda$ = 1.15 and 2.45 \AA, respectively, in the temperature range from 67 mK to 2 K. Refinements of the nuclear and magnetic structures were conducted using the {\footnotesize$\rm{FULLPROF}$} program suite~\cite{rodriguez-carvajal1993}.

\section{Results and Discussions}

\begin{figure}[htbp]
    \includegraphics[width = 8.5cm]{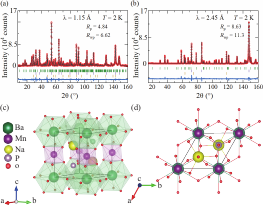}
  \caption{NPD patterns of NBMP collected 2 K using the wavelengths of 1.15 \AA~(a) and 2.45 \AA~(b), respectively. The red open circles represent the observed intensities, and the calculated patterns according to the refinements are shown as black solid lines. The differences between the observed and calculated intensities are plotted at the bottom as blue solid lines. The olive, purple, and orange vertical bars indicate the nuclear Bragg reflections from NBMP, copper cell and aluminum thermal shielding, respectively. The crystal structure of NBMP is illustrated in (c), whose projection along the $c$ axis is shown in (d).
  }
  \label{Fig1}
      \end{figure}
\vspace{1em}
\noindent 

Figures~\ref{Fig1}{(a)} and~\ref{Fig1}{(b)} show the NPD patterns of NBMP collected at 2 K (in the paramagnetic state well above $T\rm_{N2}$) using the neutron wavelengths of $\lambda$ = 1.15 and 2.45 \AA~, respectively. Both of them can be well fitted by the crystal structure with the trigonal symmetry and space group of $\mathit{P}$$\bar{3}$, the same as the room-temperature case \cite{zhang2024a}. As depicted in Figs.~\ref{Fig1}{(c)} and~\ref{Fig1}{(d)}, the Mn$^{2+}$ ions form equilateral triangular-lattice layers separated by layers of BaO$_{12}$ polyhedra, while the MnO$_{6}$ octahedra and PO$_{4}$ tetrahedra rotate in opposite directions along the $c$ axis, eliminating the mirror symmetry. No additional phases besides the sample cell and shielding can be identified from the NPD patterns, indicating high purity of our polycrystalline sample.

\begin{figure}[htbp]
    \centering
    \includegraphics[width = 8.5cm]{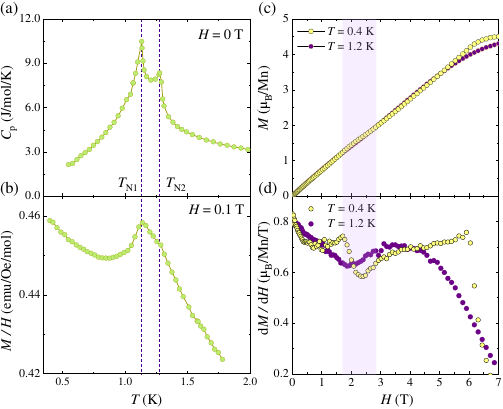}
    \caption{The low-temperature specific heat (a) and dc magnetization (b) of polycrystalline NBMP below 2 K, measured in zero field and in an applied field of 0.1 T, respectively. The vertical dashed lines mark the two transitions at $T\rm _{N1}$ and $T\rm _{N2}$. (c) and (d) shows the isothermal magnetization curves measured at 0.4 and 1.2 K and its derivative, where the shaded zone marks the region with the slope change for the $M(H)$ curve and the dip for d$M$/d$H$.}
  \label{Fig2}
      \end{figure}

The specific heat ($C\rm_p$) and dc magnetization of the polycrystalline sample of NBMP, as functions of temperature below 2 K, are shown in Figs.~\ref{Fig2}{(a)} and~\ref{Fig2}{(b)}, respectively. In the specific heat data in Fig.~\ref{Fig2}{(a)}, two sharp anomalies can be observed at $T\rm _{N1}$ $\sim$ 1.13 K  at $T_{\rm N2}$ $\sim$ 1.28 K, respectively, well consistent with those values reported in the previous study on single-crystal sample ~\cite{kim2022}. In Fig.~\ref{Fig2}{(b)},  clear responses of the magnetization at these two temperatures undoubtedly suggest the magnetic origin of both anomalies. Considering the negative Curie-Weiss temperature deduced from the paramagnetic susceptibility data as reported in Ref. [\onlinecite{zhang2024a}] and the spin-5/2 characteristic of Mn$^{2+}$ ions with a $3d^{5}$ electronic configuration, NBMP is clearly a good example of high-spin triangular-lattice antiferromagnet exhibiting two successive magnetic ordering. 

Furthermore, isothermal magnetization $M(H)$ curves of polycrystalline NBMP measured at 0.4 K (well below $T\rm _{N1}$) and 1.2 K (between $T\rm _{N1}$ and $T\rm _{N2}$) are shown in Fig.~\ref{Fig2}{(c)}, where the small initial slopes strongly suggest antiferromagnetic nature of both magnetic phases. In addition, at 0.4 K, a slope change of the $M(H)$ curve can be observed for 1.71 T $\leq$ $H$ $\leq$ 2.73 T, evidenced by a dip in its derivative (d$M$/d$H$) as shown in Fig.~\ref{Fig2}{(d)}. Such a slope change can be regarded as the finite-temperature remnant of the zero-temperature one-third magnetization plateau of the field-induced UUD quantum state, which was observed more clearly in the single-crystal sample \cite{kim2022}. 
\begin{figure*}[htbp]
  \includegraphics[width = 11.5cm]{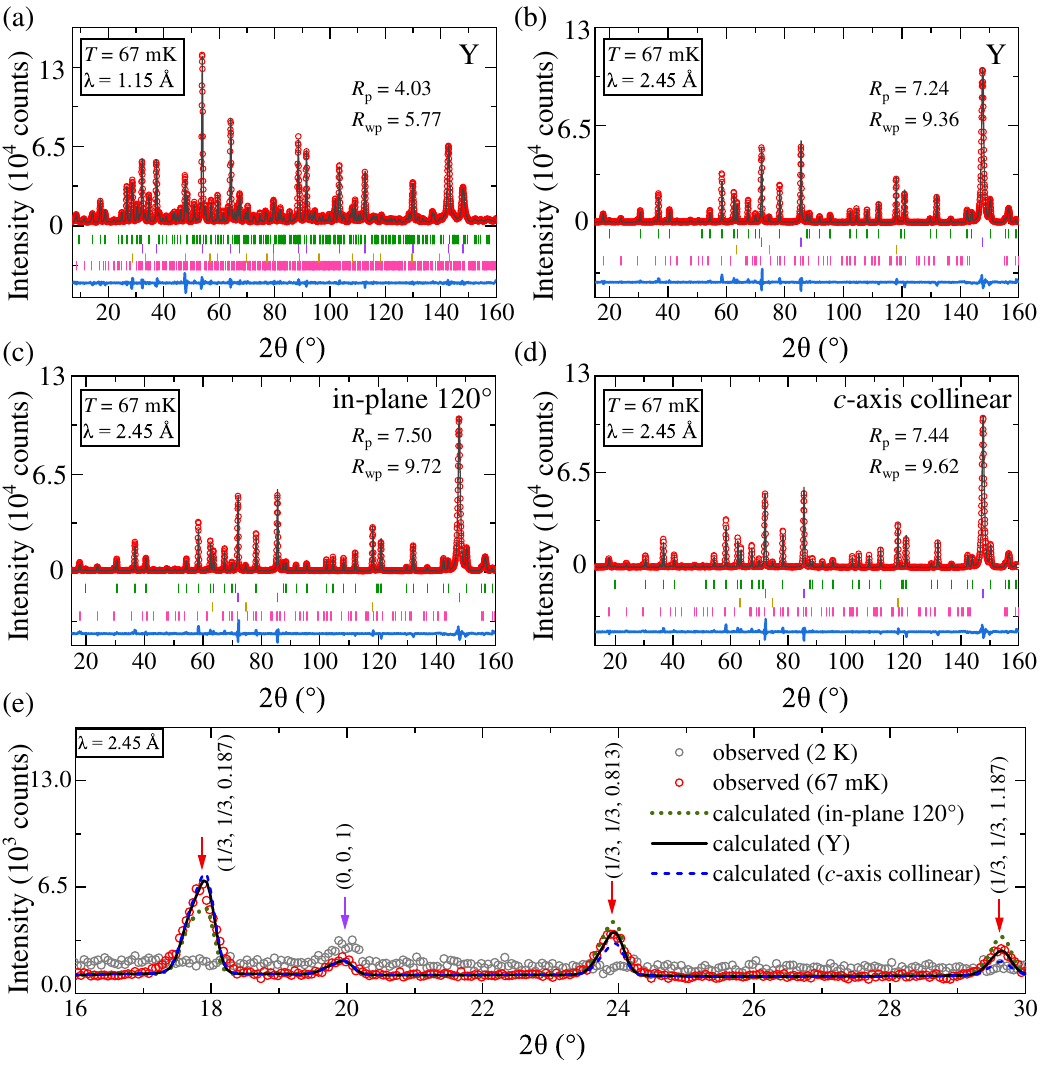}
  \caption{NPD patterns of NBMP at 67 mK and the refinements using different magnetic structure models. (a) and (b) show the data collected with the neutron wavelengths of 1.15 and 2.45 \AA, respectively, fitted using a Y-like spin structure. (c) and (d) show the fittings to the 2.45 \AA  $~$pattern using an in-plane 120$^{\circ}$ structure and a $c$-axis collinear structure, respectively. The red open circles represent the observed intensities, and the calculated patterns according to the refinements are shown as black solid lines. The differences between the observed and calculated intensities are plotted at the bottom as blue solid lines. The olive, purple, orange, and pink vertical bars indicate the nuclear reflections from NBMP, copper cell, aluminum shielding, and the magnetic reflections from NBMP, respectively. (e) shows the comparison between the refinements to the low-$Q$ data at 67 mK refined using three different models, together with 2 K data. The red vertical arrows mark the magnetic reflections, while the violet one marks the (0, 0, 1) nuclear peak.
  }
 \label{Fig3}
    \end{figure*}

Temperature-dependent NPD measurements were carried out to study the nature of the two magnetic transitions in NBMP. First, at the base temperature of 67 mK, a set of additional peaks besides the nuclear reflections observed at 2 K emerge, as marked by the pink vertical bars in Figs.~\ref{Fig3}{(a)} and~\ref{Fig3}{(b)}, which can be ascribed to antiferromagnetic (AFM) ordering of the Mn$^{2+}$ moments below $T\rm _{N1}$ and well indexed by the magnetic propagation vector of $k$ = (1/3, 1/3, 0.1869(3)). Here we note that the widths of these magnetic peaks are pretty close to the nuclear reflections, i.e., almost resolution-limited, indicating the three-dimensional (3D) nature of the magnetic order below $T\rm _{N1}$. 

The irreducible representation analysis was performed using the {\footnotesize$\rm{BASIREPS}$} program integrated into the {\footnotesize$\rm{FULLPROF}$} suite, to deduce the underlying ground-state magnetic structure. For the $\mathit{P}$$\bar{3}$ space group, the magnetic representation $\Gamma\rm_{mag}$ of the Mn$^{2+}$ ions occupying the 1b Wyckoff position of the unit cell can be decomposed as the sum of three irreducible representations (IRs), 
$\Gamma_{\mathrm{mag}}=\Gamma_{1}^{1}\oplus\Gamma_{2}^{1}\oplus\Gamma_{3}^{1}$,
whose basis vectors are listed in Table~\ref{Table1}. For $\Gamma _1$, all spins are aligned along the $c$ axis and form a $c$-axis collinear structure, but the moment size is cosinusoidally modulated along the $c$ axis due to the incommensurate $k_z$. For either $\Gamma _2$ or $\Gamma _3$, the spins form a coplanar 120$^{\circ}$ structure with opposite in-plane chirality for $\Gamma _2$ and $\Gamma _3$, but the moment direction in adjacent magnetic layers rotates spirally around the $c$ axis due to the incommensurate $k_z$. In addition to these two simple magnetic structures described by a single IR, a Y-like spin configuration, as the ground state of easy-axis triangular-lattice XXZ model \cite{Miyashita1985, Miyashita1986, Yamamoto2014}, can also be obtained by a superimposition of $\Gamma _1$, $\Gamma _2$ and $\Gamma _3$ , by fixing the ratio between the coefficients of $\Gamma _2$ and $\Gamma _3$ as 1: $-$1.

\begin{table}[]
    \caption{Basis vectors of decomposed irreducible representations (IRs) for the Mn$^{2+}$ ions locating at the 1b Wyckoff position, with a magnetic propagation vector of $k$ = (1/3, 1/3, 0.187). } 
    \begin{ruledtabular}
          \resizebox{\linewidth }{!}{  
    \begin{tabular}{ccccccc}
      ~~~~ IR ~~~ & ~~~~$m_a$ ~~~ &  ~~~~$m_b$ ~~~ &  ~~~~$m_c$ ~~~ &  ~~~~$im_a$ ~~~&  ~~~~$im_b$ ~~~&  ~~~~$im_c$ ~~~   \\ 
    \hline
    $\rm \Gamma_1$  & 0 & 0 & 1 & 0 & 0 & 0  \\
    $\rm \Gamma_2$  & 1 & 0 & 0 & $-$0.5774 & $-$1.1547 & 0  \\
    $\rm \Gamma_3$  & 1 & 0 & 0 & 0.5774 & 1.1547 & 0  \\                 
    \end{tabular}
  }
      \end{ruledtabular}
  \label{Table1}
    \end{table}

\begin{figure*}[htb]
  \includegraphics[width = 11.5cm]{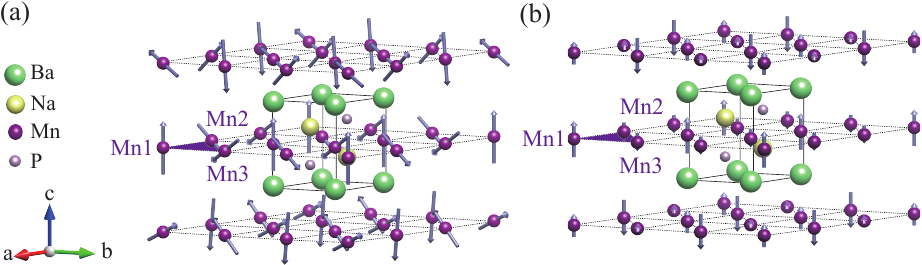}
  \caption{The Y-like (a) and the $c$-axis collinear (b) spin structure of NBMP deduced at 67 mK and 1.25 K, respectively. The colored triangle illustrates the smallest Mn1-Mn2-Mn3 triangular unit on the layer of $z$ = 0.5. The boundaries of the chemical unit cell are marked out by the solid lines.
  }
  \label{Fig4}
    \end{figure*}

Figure~\ref{Fig3} shows the comparison between the refinements to the NPD patterns at 67 mK using different possible magnetic structure models, including the Y-like structure (a), (b), in-plane 120$^{\circ}$ structure (c), and $c$-axis collinear structure (d). The difference can be better visualized in the high-resolution patterns collected using the incident neutron wavelength of $\lambda$ = 2.45 \AA. As shown in Fig.~\ref{Fig3}(e), if we inspect the strongest magnetic reflections at low 2$\theta$ angles, it is found that the in-plane 120$^{\circ}$ structure significantly underestimates the intensity of the (1/3, 1/3, 0.187) reflection around 17.9$^{\circ}$ and overestimates the intensities of the (1/3, 1/3, 0.813) and (1/3, 1/3, 1.187) reflections around 23.9$^{\circ}$ and 29.7$^{\circ}$, respectively. On the other hand, the $c$-axis collinear structure dramatically underestimates the intensities of the latter two reflections. Therefore, we conclude that either the in-plane 120$^{\circ}$ structure or the $c$-axis collinear structure described by a single basis vector of the IR is insufficient to reproduce the correct magnetic diffraction intensities and the combination of these basis vectors has to be included, yielding a Y-like ground state as shown in Fig.~\ref{Fig4}(a). The validity of this Y state is also evidenced by smaller $R$ factors of the refinement, compared with the other two model candidates, as listed in Figs.~\ref{Fig3}(b),~\ref{Fig3}(c), and~\ref{Fig3}(d). 

It is worth noting that the Y-like ground state of NBMP below $T\rm_{N1}$ $\approx$ 1.13 K is in stark contrast to the recent observation of a collinear stripy order in spin-5/2 Na$_3$Fe(PO$_4$)$_2$ below a much larger $T\rm_{N}$ of $\sim$ 10.4 K \cite{Sebastian2022,Saha2024}, a compound with a similar glaserite-type structure. The distinct magnetic properties of these two spin-5/2 compounds are largely related with their structural difference. NBMP is trigonal and Mn$^{2+}$ ions form perfect triangular layers, but Na$_3$Fe(PO$_4$)$_2$ is monoclinic and Fe$^{3+}$ ions constitute isosceles triangular lattices, in which the geometrical frustration is significantly reduced and the in-plane anisotropy favors a stripy order.

\begin{figure*}[htbp]
    \includegraphics[width = 11.5cm]{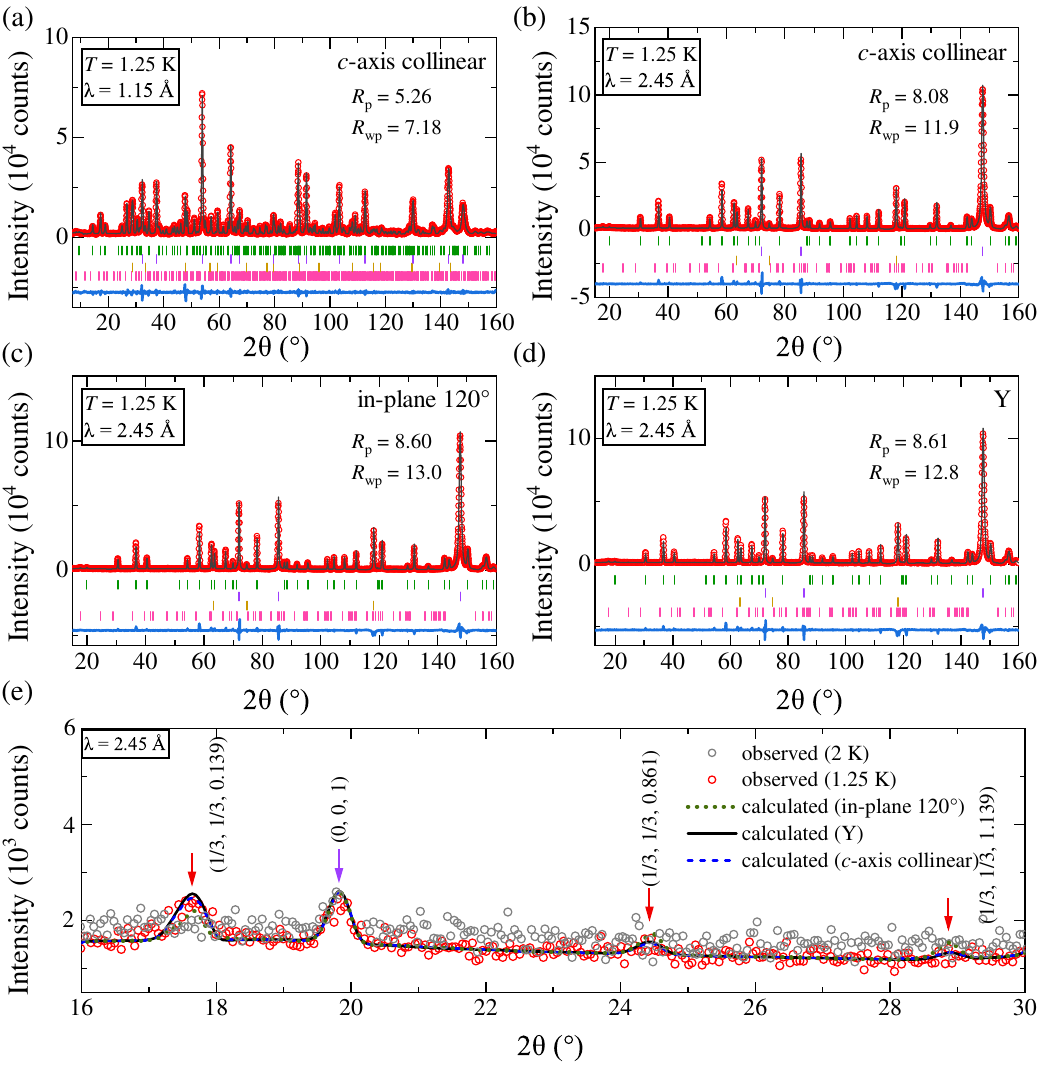}
    \caption{NPD patterns of NBMP at 1.25 K and the refinements using different magnetic structure models. (a) and (b) show the data collected with the neutron wavelengths of 1.15 and 2.45 \AA, respectively, fitted using a $c$-axis collinear spin structure. (c) and (d) show the fittings to the 2.45 \AA $~$pattern using an in-plane 120$^{\circ}$ structure and a Y-like structure, respectively. The red open circles represent the observed intensities, and the calculated patterns according to the refinements are shown as black solid lines. The differences between the observed and calculated intensities are plotted at the bottom as blue solid lines. The olive, purple, orange and pink vertical bars indicate the nuclear reflections from NBMP, copper cell, aluminum shielding, and the magnetic reflections from NBMP, respectively. (e) shows the comparison between the refinements to the low-$Q$ data at 1.25 K refined using three different models, together with 2 K data. The red vertical arrows mark the magnetic reflections, while the violet one marks the (0, 0, 1) nuclear peak.
    }
    \label{Fig5}
      \end{figure*}

\begin{figure}[htbp]
  \includegraphics[width = 8.5cm]{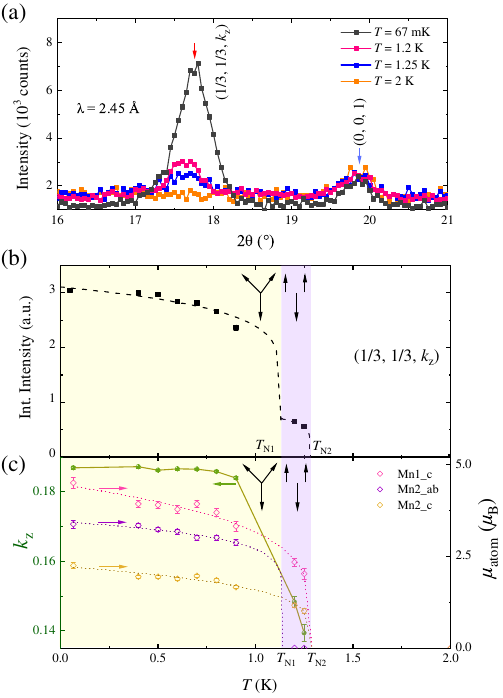}
  \caption{(a) shows the comparison between high-resolution NPD patterns at low angles collected at different temperatures using $\lambda$ = 2.45 \AA. (b) presents the temperature dependence of the integrated intensity of the strongest (1/3, 1/3, $k_z$) reflection. (c) shows the temperature dependences of the estimated out-of-plane and in-plane moment sizes (open diamonds) of the Mn1 and Mn2 ions located in the triangular unit on the layer of $z$ = 0.5 marked in Fig.~\ref{Fig4}, as well as the evolution of $k_z$ (filled spheres), the out-of-plane component of the magnetic propagation vector $k$. The dashed lines in (b) and (c) are guides to the eye, and the yellow and purple colors mark the Y and $c$-axis collinear phases for $T$  $\textless$ $T\rm_{N1}$ and $T\rm_{N1}$ $\textless$ $T$  $\textless$ $T\rm_{N2}$, respectively.
  }
  \label{Fig6}
    \end{figure}

Upon warming, the NPD patterns were also collected using two wavelengths at 1.25 K, an intermediate temperature in between $T\rm _{N1}$ and $T\rm _{N2}$, and shown in Figs.~\ref{Fig5}{(a)} and~\ref{Fig5}{(b)}. Unlike the 2 K case, magnetic diffractions due to AFM ordering can still be discerned at 1.25 K, especially around 2$\theta$ = 17.6$^\circ$, as shown in Fig.~\ref{Fig5}{(e)}, but indexed by a different propagation vector of $k$ = (1/3, 1/3, 0.139(1)). As shown in Fig.~\ref{Fig6}(c) below, the value of $k_z$, the $c$-axis component of $k$, evolves smoothly below $T\rm _{N1}$ but exhibits an abrupt jump when entering into the temperature range in between $T\rm _{N1}$ and $T\rm _{N2}$, strongly hinting the evolution into a distinct spin structure from the ground state. An irreducible representation analysis using this different $k$ yields the same basis vectors as the ground-state case (Table~\ref{Table1}). Figures~\ref{Fig5}{(b)},~\ref{Fig5}{(c)}, and~\ref{Fig5}{(d)} shows the refinements to the high-resolution pattern ($\lambda$ = 2.45 \AA) using the magnetic structure model of the $c$-axis collinear structure, in-plane 120$^{\circ}$ structure, and Y-like structure, respectively. The comparison between the fittings using these three models in the low-$Q$ range is shown in Fig.~\ref{Fig5}{(e)}. Obviously, the in-plane 120$^{\circ}$ structure underestimates the intensity of the (1/3, 1/3, 0.139) reflection and overestimates that of the (1/3, 1/3, 1.139) reflection. However, the difference between the $c$-axis collinear structure composed of one IR only and the Y-like structure composed of three IRs is quite marginal. This comparison indicate that the in-plane component does not need to be included into the magnetic structure model at 1.25 K actually, as the Y-like structure even with a lower symmetry (spin rotational U(1) symmetry broken) does not improve the fitting to the magnetic intensities at all. 
Thus, we assign the higher-symmetry $c$-axis collinear structure (with U(1) symmetry preserved) as the magnetic structure in the intermediate phase for $T\rm_{N1}$ $\textless$ $T$  $\textless$ $T\rm_{N2}$, whose spin configuration is illustrated in Fig.~\ref{Fig4}{(b)}. 

Here we need to address that this zero-field $c$-axis collinear state in the intermediate phase between $T\rm _{N1}$ and $T\rm _{N2}$ is different from the field-induced UUD state, which gives rise to the 1/3 magnetization plateau. The former is stabilized by thermal fluctuations, while the latter is stabilized by quantum fluctuations. The field-induced UUD state in triangular-lattice antiferromagnets is a ``two-$k$'' ferrimagnetic structure with a net magnetization, with two propagation vectors of $k$ = (1/3, 1/3, $k_z$) and $k$ = 0 \cite{Bordelon2019}. In our NPD pattern at 1.25 K, no ferromagnetic contributions with $k$ = 0 can be identified in addition to the antiferromagnetic diffractions with $k$ = (1/3, 1/3, 0.139), excluding the possibility of ferrimagnetism. In contrast, the $c$-axis collinear state is an antiferromagnetic structure with no net magnetization, as the moments of two $\uparrow$ spins cancel out one $\downarrow$ spin [see the magnetic structure illustrated in Fig.~\ref{Fig4}{(b)}], which is evidenced from the small initial slope of the $M(H)$ curve at 1.2 K. Such a state was theoretically expected as the zero-field intermediate phase of the antiferromagnetic triangular-lattice Heisenberg-Ising model with an easy-axis anisotropy \cite{Miyashita1985,Miyashita1986}, due to orderings of the out-of-plane components below $T\rm _{N2}$. 
Moreover, it is worth pointing out that this $c$-axis collinear structure in NBMP is an amplitude-modulated phase due to the incommensurate $k_z$, which is distinct from the non-modulated field-induced UUD structure observed in NBCP associated with the commensurate $k_z$ \cite{xiang2024}. 

Figure~\ref{Fig6}(a) summaries the high-resolution NPD patterns at low angles collected at different temperatures using $\lambda$ = 2.45 \AA. The (1/3, 1/3, $k_z$) antiferromagnetic reflection weakens with increasing temperature but persist up to 1.25 K, with its peak width always comparable to the nearby (0, 0, 1) nuclear reflection, suggesting the long-range ordered nature of the intermediate magnetic phase between $T\rm _{N1}$ and $T\rm _{N2}$. Figure~\ref{Fig6}(b) shows the temperature dependence of the integrated intensity of the strongest (1/3, 1/3, $k_z$) reflection, where two step-like behaviors are expected, due to the ordering of out-of-plane moments in the $c$-axis collinear state below $T\rm _{N2}$ and that of additional in-plane moments in the Y state below $T\rm _{N1}$, respectively.  Figure~\ref{Fig6}(c) also shows the temperature dependences of the refined out-of-plane and in-plane moment sizes of the three Mn$^{2+}$ ions located in the smallest triangular unit on the layer of $z$ = 0.5, as highlighted in Fig.~\ref{Fig4}{(a)}. For the Y-like AFM state ($T$  $\textless$ $T\rm_{N1}$), the spin of Mn1 points along the $c$ axis, while Mn2 and Mn3 have both $ab$-plane and $c$-axis spin components.  For the $c$-axis collinear state ($T\rm_{N1}$ $\textless$ $T$  $\textless$ $T\rm_{N2}$), the spins of all Mn$^{2+}$ ions purely align along the $c$ axis. Therefore, the in-plane moment of both Mn2 and Mn3 tends to vanish at $T\rm_{N1}$ already, while the out-of-plane component of all three moments persists up to $T\rm_{N2}$. This scenario account for the two separated transition temperatures observed in the macroscopic specific heat and magnetization measurements. 

The occurrence of two successive magnetic transitions in NBMP in zero field is reminiscent of a few high-spin 2D triangular-lattice easy-axis antiferromagnets including CuCrO$_2$ \cite{Kimura2008,Frontzek2012}, GdPd$_2$Al$_3$ \cite{Kitazawa1999,Inami2009}, Rb$_4$Mn(MoO$_4$)$_3$ \cite{Ishii2011} and previous theoretical investigations on triangular-lattice anisotropic Heisenberg XXZ model \cite{Miyashita1985,Miyashita1986}, where the upper and lower transitions were well attributed to the ordering of out-of-plane and in-plane spin components, respectively. It is worth pointing out that NBMP seems to be a 3D easy-axis antiferromagnet, nevertheless, in which the interlayer couplings can not be neglected, as suggested by the incomensurate ordering vector $k_z$ and almost resolution-limited sharp magnetic Bragg peaks. The easy-axis nature most likely arises from the considerable single-ion anisotropy of the Mn$^{2+}$ ions. To firmly confirm the nature of these two successive phase transitions in NBMP and to determine the spin model Hamiltonians, further investigations on high-quality single-crystal samples are necessary.

~
\section{Conclusion}
In summary, we have performed comprehensive investigations on the magnetic ordering in the spin-5/2 triangular-lattice antiferromagnet Na$_2$BaMn(PO$_4$)$_2$, combining low-temperature specific heat, dc magnetization, and zero-field NPD measurements. It is found that NBMP undergoes two successive magnetic transitions at $T\rm_{N1}$ $\approx$ 1.13 K and $T\rm_{N2}$ $\approx$ 1.28 K, respectively. Depending on the temperature, a Y-like structure and a $c$-axis collinear structure were identified as the ground-state and intermediate-phase magnetic structure, respectively. The out-of-plane component of $k$ is incommensurate in both phases, indicating non-negligible interlayer couplings. The occurrence of two successive magnetic transitions in NBMP is consistent with an easy-axis triangular-lattice AFM model.

\begin{acknowledgments}
This work is based on the experiments performed at the Swiss Spallation Neutron Source SINQ, Paul Scherrer Institute, Villigen, Switzerland. We acknowledge the financial supports from the National Key Projects for Research and Development of China (Grant No. 2023YFA1406003), the National Natural Science Foundation of China (Grant No. 12074023), the Large Scientific Facility Open Subject of Songshan Lake (KFKT2022B05), the Fundamental Research Funds for the Central Universities in China, and the Academic Excellence Foundation of BUAA for PhD Students. This work is also supported by the Synergetic Extreme Condition User Facility.

\end{acknowledgments}

\bibliography{NBMP}



\end{document}